\documentclass[pre,preprint,showpacs,floatfix,superscriptaddress]{revtex4}
\usepackage{graphicx}
\usepackage{amsmath, amsfonts,amssymb}
\usepackage{bm}
\newcommand{\be}{\begin{equation}}
\newcommand{\ee}{\end{equation}}
\newcommand{\bea}{\begin{eqnarray}}
\newcommand{\eea}{\end{eqnarray}}

\newcommand{\avg}[1]{\left\langle #1 \right\rangle}
\newcommand{\eql}[1]{\label{eq:#1}}

\def\del{\partial}
\def\dotprod{\!\cdot\!}
\newcommand{\nn}{\nonumber}
\def\gsim{\mathrel{\lower2.5pt\vbox{\lineskip=0pt\baselineskip=0pt
          \hbox{$>$}\hbox{$\sim$}}}}
\def\lsim{\mathrel{\lower2.5pt\vbox{\lineskip=0pt\baselineskip=0pt
          \hbox{$<$}\hbox{$\sim$}}}}

\begin{document}
\title{Stretching short biopolymers by fields and forces}

\author{Yuko Hori, Ashok Prasad, and Jan\'{e} Kondev}
\email{yhori@andover.edu, ashokp@mit.edu, kondev@brandeis.edu}
\affiliation{Martin Fisher School of Physics, Brandeis University,
Mailstop 057, Waltham, MA 02454-9110, USA}

 \date{\today}

\begin{abstract}
We study the mechanical properties of semi-flexible polymers, when
the contour length of the polymer is comparable to its persistence
length.  We compute the exact average end-to-end distance and shape
of the polymer for different boundary conditions, and show that
boundary effects can lead to significant deviations from the
well-known long-polymer results. We also consider the case of
stretching a uniformly charged biopolymer by an electric field, for
which we compute the average extension and the average shape, which
is shown to be trumpet-like. Our results also apply to long
biopolymers when thermal fluctuations have been smoothed out by a
large applied field or force.
\end{abstract}
 \maketitle

The mechanical properties of semi-flexible biopolymers are important for their
biological function. For example, DNA is tightly packed in eukaryotic
chromosomes and in viruses, while actin and microtubules provide the
scaffolding and structure for most animal cells~\cite{Alberts2002}. In these
cases the length scales at which the polymer properties of these macromolecules
are of biological interest are comparable to their persistence length. Namely
DNA packing typically involves loops of diameter less than its persistence
length of $50 \, {\rm nm}$, and so does looping induced by DNA bound proteins
such as the lac repressor, while actin is present in cells in the form of
bundles and networks in which the typical length of the participating polymers
is shorter than its $15 \, {\rm \mu m}$ persistence length. Stretching short
strands of DNA is also relevant for single molecule experiments involving
tethered molecules~\cite{Perkins2004}.

Mechanical properties of semi-flexible polymers are well described
by the worm-like chain model~\cite{Bustamante1994a, Marko1995c},
which treats the polymer as a space curve with a bending energy
quadratic in the curvature. On the basis of this model, the
extension of a molecule in response to an applied force can be
calculated~\cite{Marko1995c}, and has been shown to agree with
experiments to a high level of accuracy. Typically, the method of
solution has been to map the calculation of the partition function
of the worm-like chain to solving a differential equation, either
numerically~\cite{Marko1995c}, or analytically as in the
two-dimensional case~\cite{Prasad2005}.  Since experiments have
usually probed the force response of {\it long } DNA molecules,
theoretical treatments in the past have typically taken the long
chain limit, in which case boundary conditions play no role. However
this is not appropriate for molecules whose length is of the order
of a persistence length. In fact, the discrepancy between the long
chain results and the behavior of short molecules is glaringly
obvious when we look at the force extension curves themselves. All
force extension curves for long polymers pass through the origin in
the limit as force drops to zero, but it has been known for a long
time that the short worm-like chain has a finite extension at zero
force~\cite{Landau1980}.

Since the use of short DNA strands as a ''force ruler'' is becoming routine in
force spectroscopy experiments, understanding the entropic elasticity of such
strands is of considerable experimental importance~\cite{Li2006, Perkins2004}.
For example, using the formula appropriate for long molecules while fitting
force-extension measurements done on short molecules leads to incorrect
estimates of the persistence length, that could be 2--5 times smaller than the
accepted value~\cite{Li2006}.

It is interesting to note that similar considerations apply even to
long molecules, when the molecule in question is being stretched by
large forces, assuming that the entropic limit is maintained and
structural changes to the molecule are not induced. For example, if
a long molecule is stretched while keeping the tangents at the two
ends perpendicular to the direction in which the force is exerted,
the molecule will bend away from the direction of the force at the
ends. The stored length in these bends will make the force-extension
behavior of this molecule different from when it is attached with
the tangents parallel to the direction of the applied force. It has
been proposed that such a mechanism is responsible for the observed
deviations from ideal force-extension behavior when polymers are
stretched by an AFM tip~\cite{Kulic2005}.

The short chain and the large force limit discussed above are both
characterized by a small probability of large deviations from the
straight polymer configuration. Throughout this paper we refer to
this situation as the "fluctuating rod" limit of a semiflexible
polymer. In this limit we need only take into account small,
quadratic fluctuations around the energy-minimizing configuration.
This leads to a considerable simplification of the mathematics of
the worm-like chain, and allows for some elegant analytical results,
as we demonstrate below.  Note that an alternative numerical
treatment of small-chain entropic properties is provided by
accounting for the appropriate boundary conditions and finite chain
length in the series expansion of the partition function for the
worm like chain~\cite{Li2006}.

Previously, short chains were explicitly discussed in
ref.~\cite{Wilhelm1996}, who also used a harmonic approximation to
obtain the probability distribution function of the end-to-end
vector for free polymers in a series expansion. An interesting
contribution was made by ref.~\cite{Keller2003}, who discussed the
difference in the applicability of thermodynamics to short and long
polymers, and derived a formula for stretching short polymers.
Ref.~\cite{Kessler2004}, along with Monte Carlo results, rederived
this result for stretching short polymers, and also calculated
distribution functions and the effect of walls. The effect of stored
length due to boundary conditions on long polymers stretched by
large forces was calculated by ref.~\cite{Kulic2005}, who also
considered the effect of internal loops on force-extension relations
in this regime. In this paper we recover prior results for
stretching a short polymer by an applied force, using the generating
functional method. We find this field-theoretic method particularly
well suited for calculating statistical properties of fluctuating
rods. To demonstrate this we derive a number of new formulas,
including the hitherto unsolved problem of the average extension of
a charged polymer in an electric field.

\begin{figure}
\begin{center}
\includegraphics[width=\columnwidth ,height=!]{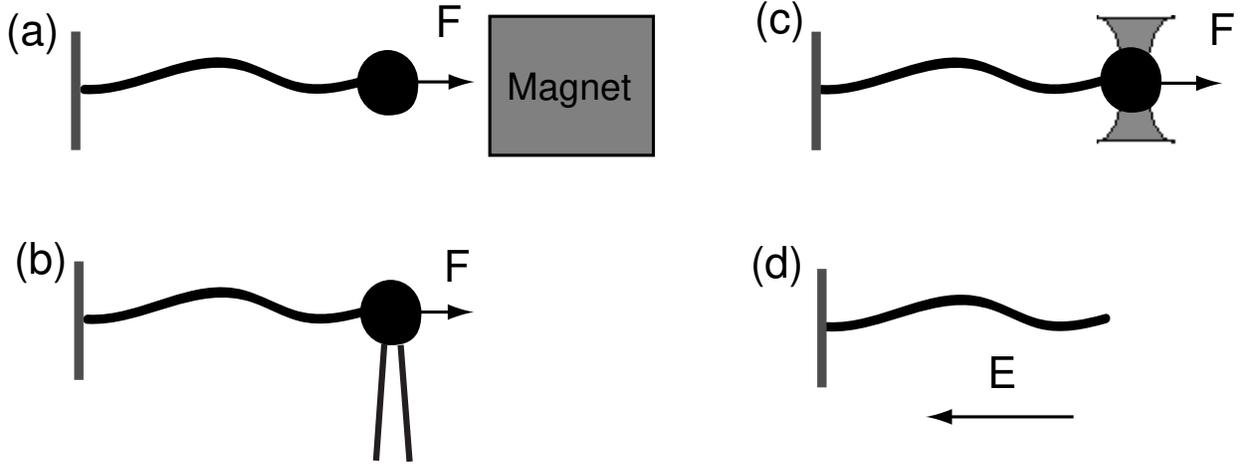}
\end{center}
\caption{\label{Setup} A cartoon of different ways to stretch single
molecules. (a) Magnetic Tweezer apparatus. The bead is paramagnetic, and the
force exerted is proportional to the spatial gradient of the magnetic field. (b) Stretching using a micropipette. The tip is imaged, and its bending is pre-calibrated with
the corresponding force exerted. (c) Laser tweezers , and (d) stretching using a uniform electric field. }
\end{figure}

This paper is organized as follows. In Sec.\ref{secII} we present the general
formalism for calculating the force-extension curves and rms-fluctuations of a
fluctuating rod under tension, using the generating functional method. In
Sec.\ref{secIII} we apply this formalism to a polymer stretched by a constant
force. However single molecule experiments that stretch semiflexible polymers
by laser tweezers, magnetic tweezers, or a micropipette~\cite{Marko1995c}
differ in the way they treat the ends of the molecule, which may be free or
constrained in several ways (Fig.\ref{Setup}). These experimental conditions
affect the entropy of the molecule, and thereby the force-extension relation.
We demonstrate this for fluctuating rods by showing that the effect of
axis-clamping at the two ends of the molecule leads to a measurable effect on
force-extension curves.

In Sec.\ref{secIV} we show that a uniformly charged polymer in a constant
electric field behaves as if it is being stretched by a force that varies
linearly along the contour. Calculating the force-extension relationship in
this case, using the standard method of obtaining the partition function by
solving the partial differential equation it satisfies, is not practical, since
the differential equation in this case is not separable. Previous work in this
area has relied on either phenomenological
arguments~\cite{Marko1995c,Maier2002}, or on linear response approximations for
weak fields~\cite{Benetatos2004}. However, the general method we outline in
Sec.\ref{secII} can easily handle this situation as well, and we present new
analytical results for the force-extension relation and the shape of a
uniformly charged fluctuating-rod polymer stretched by an electric field. The
calculated average shape is in accord with observations~\cite{Maier2002}. Our
results are applicable to stiff polymers like actin for a wide range of field
strengths, and can be used to  measure the effective charge density of actin
for different salt conditions.

\section{Statistical mechanics of fluctuating rods}
\label{secII}
Since stretching experiments can be performed in both two and three dimensions,
we work in general $d$-dimensions. The objective is to calculate the average
extension of the polymer in the direction of the applied force or field. We
start by making the small-fluctuation approximation, which transforms the
partition function of the worm-like-chain into a Gaussian path integral. Using
this partition function we obtain the tangent-tangent correlation function,
from which we derive the average extension of the molecule.

The Hamiltonian of a worm-like-chain polymer of contour length $L$, which is
stretched by a constant force $\bm{F}$ is, \be \beta H[{\bm t}(s)]=\int^L_0
\!\!\! d s \left\{\frac{\xi}{2} \left( \frac{d {\bm t}}{d s} \right)^2
 -\beta{\bm F}\dotprod {\bm t}(s)\right\}~,
\label{hamil} \ee where $\beta=1/k_BT$, $s$ is the arc length and
$\xi$ is the bending stiffness and is simply related to the
persistence length, $l_p= 2 \xi /(d-1)$, in $d$-dimensions. For
double-stranded, DNA $\xi \sim 50 \, {\rm nm}$ and for an actin
filament, $\xi \sim 15 \, {\rm \mu m}$. The unit tangent vector
${\bm t}(s)$ specifies the conformation of the chain. The polymer is
stretched by tethering the $s=0$ end to a fixed support, usually a
bead held by a micropipette, and pulling on the other end. The usual
procedure is to attach either a magnetic or a polystyrene bead to
the other end of the polymer, and to exert a constant force ${\bm F}
$ using magnetic tweezers, optical tweezers, or another micropipette
(Fig.\ref{Setup}). We assume that ${\bm F}$ is along the $\bm{e}_1 $
direction in $d $-dimensional space. The components of the tangent
vector are ${\bm t}(s)=\{\pm \sqrt{1- \sum^{d}_{i=2} t_i(s)^2}, \,
t_2(s),\cdots, t_{d}(s)\}$. In the limit of small fluctuations, the
tangent vector makes only very small deviations away from the
direction of the applied force, so that its coordinate along the
$\bm{e}_1$ direction can be approximated as, \be t_1(s) = 1-\frac12
\sum^{d}_{i=2} t_i(s)^2 ~. \label{smallfluc} \ee In this limit,
$(d{\bm t}/ds)^2 \approx \sum^{d}_{i=2}(dt_i/ds)^2$. Inserting these
expressions in the Hamiltonian, Eq.(\ref{hamil}), and integrating
the bending energy term by parts yields, \be \beta H[{t_i(s)}] =
\frac12  \sum^{d}_{i=2} \left[\int^L_0 \!\!\! d s t_i(s){\cal
O}(s)t_i(s) + B_i(L,0) \right] ~, \label{hquadratic} \ee where we
have left out an unimportant constant term. Here the term
$B_i(L,0)\equiv \frac{\xi}{2} t_i(s) \frac{d t_i}{d s}\mid^L_0$
depends on the boundary conditions, while \be {\cal O}(s)\equiv -\xi
\frac{d^2}{ds^2} + \beta F(s) \label{bigO} \ee is a differential
operator. We consider the general case of an $s$-dependent force,
$\bm{F}(s)$, which includes the case of stretching by an electric
field discussed in Sec.\ref{secIV}.

The boundary term $B_i(L,0)$ depends in general on the tangent vectors at the
two ends of the polymer, and for the cases discussed here it vanishes. Namely,
the tangent vector at the end is either unconstrained, in which case $d\bm{t}/ds =0 $ for the end-tangent vector, or aligned with the force, and then $t_i=0$ for
$i=2,3 \ldots$.

The partition function of the fluctuating rod, ${\cal Z} $,  is a
path integral of the Boltzmann factor, $\exp (-\beta H[{t_i(s)}]) $,
over all possible conformations of the polymer, where $H[{t_i(s)}] $
is given by Eq.(\ref{hquadratic}) with the above boundary
conditions. To compute the tangent-tangent correlation function we
employ the generating functional, ${\cal Z}(J_i) $, which is
obtained by adding a source term, $J_i(s)$ to the Hamiltonian: \be
{\cal Z}(J_i) = \int {\cal D}[\{\bm{t}(s)\}]\exp \left[-\beta
H[\{t_i(s)\}] + J_i(s)t_i(s) \right] ~. \label{genfun} \ee The
correlation function, $\langle t_i(s) t_i(s')\rangle $ is then
obtained by taking functional derivatives~\cite{Peskin1995}, \be
\langle t_i(s) t_i(s')\rangle = \lim_{J_i\rightarrow 0}
\frac{\delta^2 \log[{\cal Z}(J_i)]}{\delta J_i(s) \delta J_i(s') } =
G(s,s') ~, \label{tstsprime} \ee where, $G(s,s') $ is the Green's
function of the operator ${\cal O}(s)$ defined by, \be {\cal
O}(s)G(s,s') = \left[ -\xi \frac{d^2}{ds^2} + \beta F(s) \right]
G(s,s')=\delta(s-s') ~. \label{Gss} \ee

Note that, by symmetry considerations, fluctuations transverse to
the applied force are equivalent for all transverse directions,
hence we need to calculate this Green's function for only one of the
$d-1$ directions.

To calculate the average extension in the $x_1$-direction at any
contour position $s$, we now make use of the relation \be
\avg{x_1(s)} = \int^s_0 \!\!\! d s' \left( 1-
\frac{d-1}{2}\avg{t_i(s')^2} \right) ~, \label{x(u)} \ee which is
valid in the small fluctuation approximation regime given by
Eq.(\ref{smallfluc}). The end-to-end extension is $X =\avg{x_1(L)}$.
The average of each of the other $d-1$ orthogonal coordinates,
$x_i\; (i=2,\ldots ,d)$, is zero, while their root-mean-square (RMS) value
is \be \avg{x_i(s)^2}^\frac12 = \left(\int^s_0 \!\!\! d s' \int^s_0
\!\!\! d s''
          \avg{t_i(s')t_i(s'')}\right)^\frac12 ~.
\label{y(u)}
\ee
A parametric plot of the mean extension against the
root-mean-square value of $x_i(s),\;(i=2,\ldots,d)$ using Eq.(\ref{x(u)}) and
Eq.(\ref{y(u)}) describes the average shape of the polymer.

An important question is, under what conditions does the fluctuating rod
approximation apply? For this we employ the tangent-tangent correlation
function which yields a self-consistency condition for the small-fluctuation
assumption. In two dimensions, the small fluctuation
approximation, Eq.(\ref{smallfluc}), implies that $t_x \approx 1 - t_y^2/2$ for
all $s$, which is accurate to within $1 \% $ when $t_y^2 < 1/2$. In
$d$-dimensions this suggests the condition that the sum of all mean-squared
fluctuations at any contour position is at most $1/2$. In other words, we
require that \be \avg{t_i(s)^2} \leq 1/(2d-2) ~,
\label{cond}
\ee for all $s$. Below we repeatedly make use of this condition in order to
estimate the range of experimental parameters (polymer length, magnitude of
force, etc.) for which the fluctuating rod model is applicable.

\section{Stretching by a constant force}
\label{secIII}

To illustrate the method described above, we calculate explicitly the
force-extension relation and the rms-fluctuations of a fluctuating rod
stretched by a force applied at one of its ends. The experimental setup is as
shown in Fig.\ref{Setup}a. The correlation function, $\langle
\bm{t}(s)\bm{t}(s')\rangle$ is given by the Green's function of the
differential operator, Eq.(\ref{bigO}). Note that the assumption that the
tangent vectors at the two ends of the polymer are both aligned with the
direction of force makes the Green's function vanish at the boundaries.

The differential equation is solved separately in the domains $s < s'$ and $s >
s'$, in which the delta function vanishes, and then the solutions are matched
at the boundary $s = s'$. This yields,
\begin{widetext}
\bea
G(s,s')=
\left\{ \begin{array}{cr}
{\cal N}\sinh [\kappa_F s] \sinh [\kappa_F(L-s')]\equiv G_{<}(s,s'), &
\text{when $s < s'$}\\
{\cal N'}\sinh [\kappa_F(L-s)] \sinh [\kappa_F s'] \equiv G_{>}(s,s'),
& \text{when $s > s'$}
\end{array} \right. ~,
\eql{Gpartial}
\eea
where
\be
\kappa_F \equiv \sqrt{\frac{F}{\xi k_B T}} ~,
\label{kappaf}
\ee is a force-dependent inverse length. Requiring the continuity of $G(s,s') $
at $s=s' $, we find $\cal{ N}=\cal{N'} $. Also, by integrating both sides of
Eq.(\ref{Gss}) over the interval $s \in (s-\epsilon , s + \epsilon)$ and taking
the limit $\varepsilon \rightarrow 0 $ gives the final boundary condition on
$G(s,s') $: \be
 \left.\frac{\del G_{>}}{\del s}\right|_{s=s'}-
\left.\frac{\del G_{<}}{\del s}\right|_{s=s'} = -\frac{1}{\xi} ~.
\label{Gbc}
\ee
Putting it all together, we get,
\bea
G(s,s')=
\left\{ \begin{array}{cr}\displaystyle
\frac{\sinh [\kappa_F s] \sinh [\kappa_F(L-s')]}{\kappa_F \xi \sinh [\kappa_F L]},
 &\text{when $s \leq s'$}\\
\displaystyle
\frac{\sinh [\kappa_F(L-s)] \sinh [\kappa_Fs']}{\kappa_F \xi \sinh [\kappa_F L]},
& \text{when $s \geq s'$}
\end{array} \right. ~.
\label{gdirichlet}
\eea

\end{widetext}

\begin{figure}
\begin{center}
\includegraphics[width=4in ,height=!]{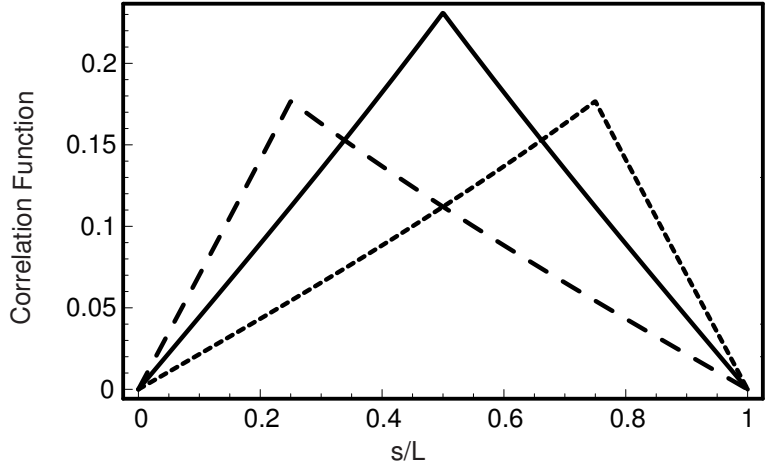}
\end{center}
\caption{\label{GSS}The Greens function $G(s,s')$, with $\kappa_F = 1$, plotted for $s'/L=0.25$ (long dashes), $0.5$, (solid line) and $0.75 $ (small dashes). }
\end{figure}

This Green's function is plotted in Fig.\ref{GSS}.

\begin{figure}
\begin{center}
\includegraphics[width=4in ,height=!]{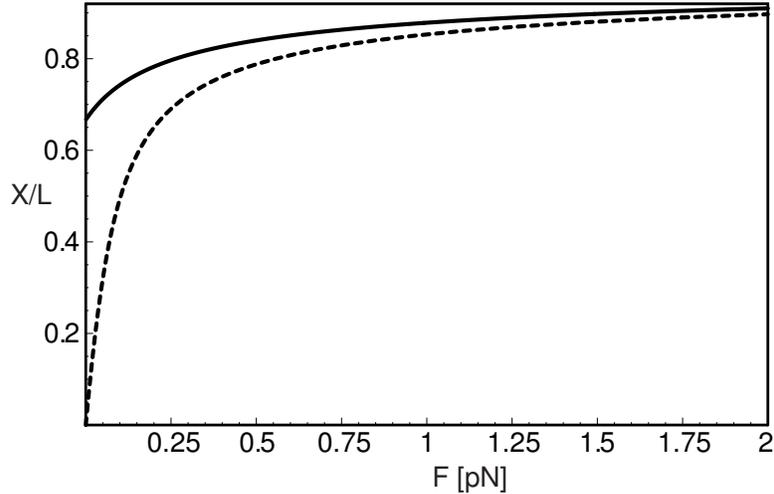}
\end{center}
\caption{\label{MagTweezer} The force-extension curve for a $100$ nm strand of
DNA, plotted against force in piconewtons (solid line) using Eq.(\ref{fextshort}).
The dashed line shows the force-extension curve for a
long polymer, plotted using the approximate interpolation formula of Ref.~\cite{Marko1995c}}
\end{figure}

When $s=s'$, the correlation function yields the mean-square fluctuations of
the tangent vector coordinate, $\langle t_i(s)^2\rangle $. Plugging this into
Eq.(\ref{x(u)}), we obtain
\be \frac{\langle x_1(s)\rangle}{L} = \frac{s}{L} -
(d-1)\frac{2\kappa_F s \cosh[\kappa_F L] -\sinh[\kappa_F L] + \sinh[\kappa_F
(L-2s)] }{8\kappa_F^2 \xi L} ~.
\label{xone(u)}
\ee
Now we set $s=L $ in Eq.(\ref{xone(u)}) to get the force extension relation
for a fluctuating rod polymer, \be \frac{X}{L} =  1 -
\frac{d-1}{2}\frac{\kappa_F L\cosh[\kappa_F L]-\sinh[\kappa_F L]} {2 \kappa_F^2
\xi L \sinh[\kappa_F L]}
\label{fextshort}
\ee A plot of this relation, and its comparison with the long-polymer result is
shown in Fig.\ref{MagTweezer}. For $d=3 $, Eq.(\ref{fextshort}) agrees with the
force-extension relation computed in Ref.~\cite{Keller2003}.

Experiments that involve stretching short strands of DNA, or other semiflexible molecules, should use
Eq.(\ref{fextshort}) in place of the force-extension relation of
Ref.~\cite{Marko1995c}, which is appropriate in the long chain limit~\cite{Khalil2007}. As we
have remarked earlier, use of the long chain formula leads to erroneous results
when fitting data from stretching short polymers~\cite{Li2006}. It is
reassuring to note that, as we shall show below, Eq.(\ref{fextshort})
reproduces known results in the limits of high force and zero force.

It should be also noted that in the long-polymer limit, the statistical
properties of a semi-flexible polymer under an applied force depend on one
dimensionless ratio, $F\xi/(k_BT)$, which also delineates the limits of high
and low force. In the case of fluctuating rods however, both of the two
independent dimensionless ratios that can be formed by the three lengths, $L,
\xi $ and $\kappa_F^{-1} $, are present in the functions describing the
statistical quantities of interest. Also, in this case the high force limit is
governed by a different dimensionless number, $\kappa_F L \equiv \sqrt{FL^2/\xi
k_BT} $.

We can also calculate the root-mean-square fluctuations of the polymer
in the transverse directions, using Eq.(\ref{y(u)}),
\be
\frac{\avg{x_i(s)^2}^{\!\frac12}}{L}  =   \left(\frac{2 \kappa_F s \sinh[\kappa_F L] - 3 \cosh[\kappa_F L]
   + 4 \cosh[\kappa_F (L-s)]- \cosh[\kappa_F (L-2s)]}
   {2 \kappa_F^3 L^2 \xi \sinh[\kappa_F L]}
     \right)^{\!\!\frac{1}{2}} ~,
\label{yfluc}
\ee
for the transverse directions $i = 2, \dots d $.
A parametric plot of $\langle x_1(s)\rangle$ given by Eq.(\ref{xone(u)})
against $\sqrt{\langle x_i(s)^2 \rangle}\ i = 2,\ldots d $ defines
the average shape of the polymer.

\begin{figure}
\begin{center}
\includegraphics[width=4in ,height=!]{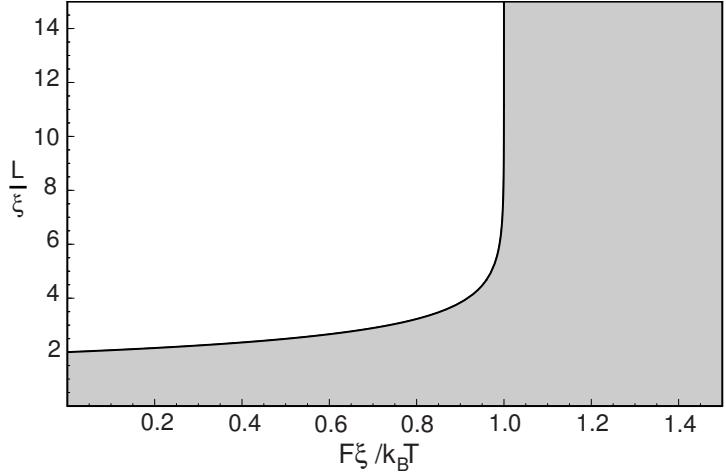}
\end{center}
\caption{\label{ConstCond} The set of parameter values that satisfy the small
fluctuation approximation is shown by the shaded area. The $x $-axis is the
reduced force, $\kappa_F^{2}\xi^2 \equiv F\xi/k_B T$. The $y$-axis is the
polymer length in terms of the $3d$-persistence length, $L/\xi$.}
\end{figure}

The self-consistency assumption, Eq.(\ref{cond}), is a necessary
condition for the validity of Eq.(\ref{fextshort}) and
Eq.(\ref{yfluc}). It is clear from Eq.(\ref{gdirichlet}) that the
maximum fluctuations of the tangent vector are at the center of the
chain, $s=L/2$. We therefore require that \be \avg{t_i(s)^2} \leq
\avg{t\left (\frac{L}{2} \right )^2} \leq \frac{1}{2(d-1)}~.
\label{condition1} \ee Taking $d=2$ for illustration, this yields
the condition that \be \displaystyle \frac{1}{2 \kappa_F \xi} \tanh
\displaystyle \!\! \left[\frac{\kappa_F L }{2}\right] \leq
\frac{1}{2} ~. \label{theta2_max} \ee As the hyperbolic tangent is
never greater than one, Eq.(\ref{theta2_max}) always holds when \be
  \kappa_F \xi  \geq 1
\label{f_only} ~.
\ee
However, even when the force is weak, Eq.(\ref{fextshort}) can apply,
provided the polymer is short enough compared to its persistence length. In
other words, even when $\frac{1}{2 \kappa_F \xi} > 1$, the small fluctuation
assumption can be satisfied, provided the hyperbolic tangent is small enough.
In this limit Eq.(\ref{theta2_max}) can be rewritten as \be \frac{L}{\xi} \leq
\frac{1}{\kappa_F \xi}\log \!\! \left[\frac{1+\kappa_F \xi}{1-\kappa_F
\xi}\right] ~.
\label{l_vs_f}
\ee Conditions (\ref{f_only}) and (\ref{l_vs_f}) are summarized graphically by
the shaded region in Fig.\ref{ConstCond}.

It is an interesting exercise to examine the limits of
Eq.(\ref{fextshort}) in the two cases above, described by
Eq.(\ref{f_only}) and Eq.(\ref{l_vs_f}). When the polymer is long,
so that $L/\xi \gg 1$, the strong force condition,
Eq.(\ref{f_only}), ensures that the product $\kappa_F L  \gg 1$.
Then the force-extension curve in Eq.(\ref{fextshort}) reduces to
\be \frac{X}{L} \simeq 1-\frac{d-1}{4} \sqrt{\frac{k_B T}{F \xi}} ~,
\label{markosiggia} \ee which, for $d=3 $, is the well known result
for high force stretching~\cite{Marko1995c}.  Our result in
Eq.(\ref{fextshort}) shows, in fact, that even for a short chain
satisfying $L/\xi \leq 4$, as long as the force is strong enough,
such that the product $\kappa_F L \gg 1$, we recover a
force-extension relation of the form Eq.(\ref{markosiggia}).

The other limit is the case when the force is weak such that
$\kappa_F L \leq 1$, then the force extension curve
in Eq.(\ref{fextshort}) becomes
\be
\frac{X}{L} \simeq
 1- \frac{d-1}{2}\left(\frac{1}{6}\left(\frac{L}{\xi}\right) - \frac{F \xi}{90
k_B T} \left(\frac{L}{\xi}\right)^3\right) ~.
\label{weak-force}
\ee
This is different from the behavior of long worm-like chain polymers under weak forces,
in which case the force-extension relation is  proportional to the force~\cite{Marko1995c},
with no constant term.
Note also that when the force goes to zero, the relative extension of the polymer tends
to $1- \frac{d-1}{2}(\frac{1}{6})\left(\frac{L}{\xi}\right)$, which is a result one finds
 in Landau and Lifshitz for $d=3 $~\cite{Landau1980}.

What about the case when $1/(2 \kappa_F \xi) > 1 $? Then we must
have $\tanh(\kappa_F L/2) < \kappa_F \xi $, or in other words, to a
first approximation, \be L < 2 \xi ~, \label{short} \ee which can be
confirmed by a look at Fig.\ref{MagTweezer}. Hence polymers up to
one persistence length in two dimensions, or two persistence lengths
in three dimensions, can be regarded as  satisfying the small
fluctuation approximation for any applied force.

\subsection{Effects of Axis Clamping}

We now show how axis clamping affects the force extension for a
fluctuating rod. If the free end of the polymer is held rigidly,
such that the end-to-end separation vector of the polymer is
constrained to be collinear with the direction of the force, there
is an additional constraint, \be \int^L_0 \!\!\! d s \,\, t_i(s) = 0
\quad \mbox{for} \quad i = 2,3,\ldots , d .\label{clampcons} \ee
Note that this constraint depends upon the entire conformation of
the chain, hence it is not \textit{a priori} clear when it can be
ignored.

\begin{figure}
\begin{center}
\includegraphics[width=4in,height=!]{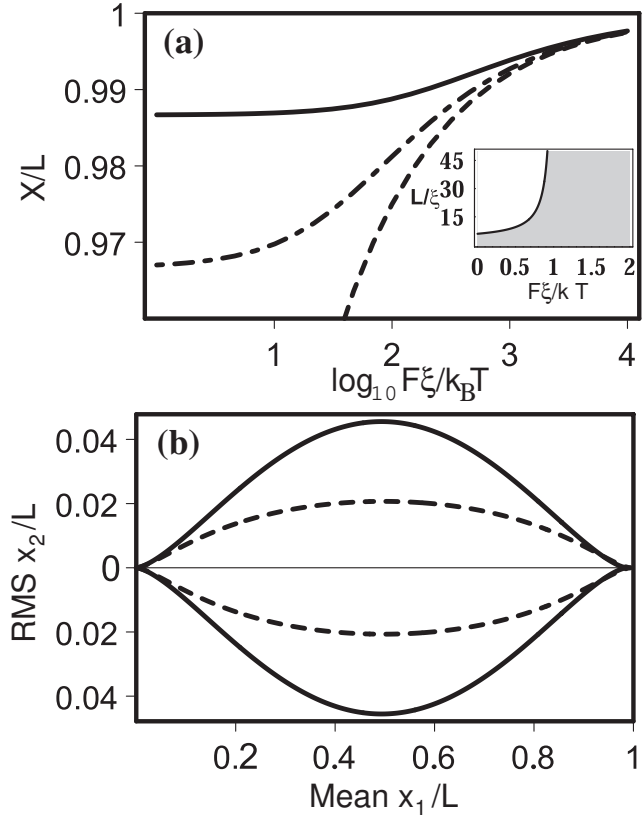}
\end{center}
\caption{(a) The relative extension of a polymer
$X/L$ pulled by a constant force in the presence of axis-clamping
is plotted against the reduced force $F \xi/(k_B T)$, for a short polymer
(solid line) for $d=2$. The dot-dashed line is the extension without the axis-clamping condition, Eq.(\ref{fextshort}) while the dashed curve is the long chain result~\cite{Prasad2005}. The reduced polymer length is
$L/\xi =0.4$ ($\approx 20 \, {\rm nm}$ for DNA and $\approx 6 {\rm \mu m}$ for
actin). In terms of force, $F \xi/(k_B T)=1$ means  $0.082 \, {\rm pN}$ for DNA, and $0.27 \, {\rm fN}$ for actin. Inset shows the range of allowed parameter values (shaded area). (b) The shape of a polymer stretched
with a laser tweezer is plotted parametrically in units of polymer contour
length $L$ for a short polymer  ($L/\xi = 0.4$). The solid curve is for $F\xi/(k_BT) = 1$, and the dotted curve is for $F\xi/(k_B T)=1000$.}
\label{LaserTweezer}
\end{figure}

We can compute the correlation function, $\avg{t_i(s)t_i(s')}$, for all $i =
2, 3, \ldots, d $, by forcing the new constraint in Eq.(\ref{clampcons}) using
the Dirac delta function,
\begin{widetext}
\be \avg{t_i(s) t_i(s')} = \frac{\displaystyle \int \!\! {\cal D} t_i(s)\>\>
t_i(s)t_i(s')
              \> \delta \! \left(\int^L_0 \!\!\! d s \> t_i(s)\right)
              \exp \! \left[\int^L_0 \!\!\! d s \left\{-\frac12 \,t_i(s)
              {\cal O}(s) t_i(s) \right\} +\beta F L \right]}
           {\displaystyle \int \!\! {\cal D} t_i(s)
           \> \delta \! \left(\int^L_0 \!\!\! d s \> t_i(s)\right)
              \exp \!\left[\int^L_0 \!\!\! d s \left\{-\frac{1}{2}\, t_i(s)
              {\cal O}(s) t_i(s) \right\} +\beta F L \right]}
\label{clampcorr}
\ee
\end{widetext}
We use the integral representation of the delta function in Fourier
space, add the source terms as before and compute the Gaussian
integrals by completing the square. This gives us an additional term
in the correlation function, \be \avg{t_i(s)t_i (s')}  =  G(s,s')  -
 \!\!  \frac{\displaystyle
        \left(\int^L_0 \!\! d s_1 G(s,s_1)\right)\!\!
          \left(\int^L_0 \!\! d s_2 G(s',s_2)\right)}
           {\displaystyle \int^L_0 \!\! d s_1
              \!\! \int^L_0 \!\! d s_2 G(s_1,s_2)}   ~,
\label{tcorrlaser} \ee where $G(s,s')$ is the Green's function of
the operator ${\cal O}(s)$ shown in Eq.(\ref{gdirichlet}). Now using
Eq.(\ref{tcorrlaser}) in Eq.(\ref{x(u)}), we obtain the average
extension, \bea \frac{X}{L} =  1 - \frac{(d-1)(\kappa_F
L\cosh[\kappa_F L]-\sinh[\kappa_F L])}
{4\kappa_F^2 \xi L \sinh[\kappa_F L]} & + &\nn \\
 \frac{(d-1)(\kappa_F L\cosh[\kappa_F L]-3\sinh[\kappa_F L]
+ 2\kappa_F L)}{2\kappa_F^3\xi L^2 + 2\kappa_F^3\xi L^2 \cosh[\kappa_F L]- 4 \kappa_F^2 \xi L\sinh[\kappa_F L]} &.&
\label{extlaser}
\eea The first two terms in Eq.(\ref{extlaser}) are identical to
Eq.(\ref{fextshort}) derived above.  Therefore, we find that the loss in
entropy due to axis clamping of the two ends of the polymer, leads to a small
but measurable correction to the average extension. This is plotted in
Fig.\ref{LaserTweezer} for $d=2$. We can also use Eq.(\ref{y(u)}) and calculate
the average shape of the polymer, defined by Eq.(\ref{x(u)}) and
Eq.(\ref{y(u)}), by plotting $\avg{x_1(s)}$ against $\sqrt{\avg{x_2(s)^2}}$
parametrically, as shown in Fig.\ref{LaserTweezer}b.

We calculate the range of validity of Eq.(\ref{extlaser}) by imposing Eq.(\ref{cond}) as before. The parameter values that satisfy this condition are
displayed in the inset of Fig.\ref{LaserTweezer}. It can be shown that when
$L/\xi \leq 6 $, Eq.(\ref{extlaser}) holds for all $F$, while for long polymers
axis clamping has no effect on the force-extension relationship, and for
$\kappa_F L\gg 1$ we again recover the well-known result~\cite{Marko1995c},
Eq.(\ref{markosiggia}).

We can expand Eq.(\ref{extlaser}) in the limit of small forces to
obtain an equation analogous to Eq.(\ref{weak-force}) for the case
of axis clamping. Interestingly, for $d=3 $ we obtain, \be
\frac{X}{L} = 1 - \frac{1}{30}\left(\frac{L}{\xi}\right) + \frac{11
F \xi}{25200k_BT}\left(\frac{L}{\xi}\right)^3 ~.
\label{weak-force-laser} \ee Note that as $F\rightarrow 0$, $X/L $
approaches $1-L/(30 \xi) $, while  Eq.(\ref{weak-force}) shows that
without the entropic constraint, $X/L$ approaches $1 - L/(6\xi ) $
in the limit of zero force. Hence we see a  different behavior at
low forces due to the entropic constraint at the end.

An experimentally relevant situation is illustrated in
Fig.~\ref{Setup}b, when the polymer is stretched using a
micropipette and the force exerted is calculated by the bending of
the micropipette tip. An equivalent setup uses a bead attached to a glass fiber and measures the bending of the fiber. These methods have been used earlier, for example to stretch chromosomes~\cite{Houchmandzadeh1997} and long DNA molecules~\cite{Cluzel1996}. Stretching using an AFM also involves axis
clamping, though here the applicability of our formulae depends on
the other boundary condition, i.e.~the collinearity of the first and
last tangent vectors with the force, being satisfied. In principle,
the entropic effect of axis-clamping may also be observed in
stretching with a laser tweezer since the optical bead is held in a
three dimensional trap formed by the laser beam (Fig.\ref{Setup}c).
In practice however the trap is not a perfect clamp. First, it is
significantly weaker in the $z$-direction, so some fluctuations do
occur. Second, it is very hard to control where on the surface of
the bead the polymer binds, so collinearity may be only
approximately satisfied, since the bead is about three orders of
magnitude larger that the polymer thickness. Finally, the bead is
free to rotate in the trap, to the extent allowed by the twist
elasticity of the polymer~\cite{Ref06}. Experimentally the effect of
axis clamping might be relevant for stretching very short DNA
strands -- a $0.1 \rm{pN}$ force will stretch a $300 \rm{nm}$ strand
of DNA to $0.86$ of its contour length if stretched with
axis-clamping, compared to $0.81$ without it. The difference might
also be observable when pulling on actin of about a persistence
length in size, but in the range of femtonewton forces.

\section{Fluctuating rod in an electric field}
\label{secIV}
Next we consider stretching of a charged polymer by a constant electric field.
Here, one end of the polymer is tethered, and when the field is switched on,
the polymer is observed to extend~\cite{Maier2002,Ferree2003}. Because the
polymer is in a charged aqueous environment, the molecular mechanism of the
electrophoretic stretch is complicated~\cite{Stigter1998,Ferree2003}. Here we
abstract from these difficulties and assume that the polymer responds to the
field ${\bm E}$ as a uniformly charged rod with charge per unit length
$\lambda$.

As in the previous section the contour length along the polymer is denoted by
$s$, with $s=0 $ denoting the tethered end. If the position vector of the
segment $s $ along the contour is denoted by ${\bm r}(s)$, the interaction
potential of the polymer with the field can be written as, \be H_I[{\bm r}(s)]
= - \lambda {\bm E}\dotprod \int^L_0 ds
   \left( {\bm r}(s) - {\bf r}(0) \right) ~.
\label{h-i}
\ee
Note that we are not only assuming that the effective charge density is
constant, but that it also remains unchanged during the course of stretching. We
can express Eq.(\ref{h-i}) in terms of tangent vectors,
\be
H_I[{\bm t}(s)]= -\lambda {\bm E}\cdot  \int^L_0 ds  \int^s_0 ds' {\bm t}(s')
\label{hint}
\ee

We now change the order of integration. Instead of integrating first over $s'$
from $0 $ to $s $, and then over $s $ from $0 $ to $L $, we can equivalently
first integrate over $s $, from $s' $ to $L $, then integrate over $s' $, from
$0$ to $L$. Hence we get,
\begin{eqnarray}
H_I&=&  -\lambda {\bm E}\cdot  \int^L_0 ds'  \int^L_{s'} ds {\bm t}(s') \nn \\
&=&  -\int^L_0 ds'  \lambda {\bm E}(L-s')\cdot {\bm t}(s')
\label{h-i-t} ~,
\end{eqnarray}
Thus a constant electric field stretches on a uniformly charged polymer as
though it were subjected to a contour dependent force, ${\bm F}(s)\equiv {\bm
E}\lambda (L-s)$. This force is zero at the free end, $s=L$, and reaches a
maximum of $|\bm{E}|\lambda L$ at the tethered end.

The $s$-dependent potential $H_I$ makes the application of spectral
methods to evaluating the partition function not practical. Namely,
in this case the calculation of the eigenfunctions for the spectral
representation of the partition function maps to a Schrodinger-like
equation with a time dependent potential, making the analysis quite
complicated. The generating functional formalism, on the other hand,
can be employed without much difficulty.

The boundary conditions appropriate for this situation are $t_i(0)=0,\
i=2,3,\dots , d$ at the tethered end, and $\frac{dt_i(s)}{ds}|_{s=L}=0, \
\mbox{for all}\  i $ at the free end. The boundary term in
Eq.(\ref{hquadratic}) once again vanishes. The correlation function
$\langle\theta_i(s)\theta_i(s')\rangle$ is now the Greens function, $G(s,s')$,
of the operator \be {\cal O}(s) \equiv -\xi \frac{d^2}{d s^2}+\beta \lambda E
(L-s) ~,
\label{efieldop}
\ee
with the boundary conditions above.

The solution to
Eq.({\ref{Gss}) that satisfies the boundary conditions is calculated in the
same way as for the constant force case discussed above, but now in terms of Airy functions,
$Ai$ and $Bi$, and their derivatives, $Ai' $ and $Bi' $,
\begin{widetext}
\be
G(s,s') = \frac{ \pi \left({\rm Ai}[\kappa_E(L-s)]{\rm Bi'}[0]
-{\rm Ai'}[0]{\rm Bi}[\kappa_E(L-s)]\right)
\left({\rm Ai}[\kappa_E(L-s')]{\rm Bi}[\kappa_E L]
-{\rm Ai}[\kappa_E L]{\rm Bi}[\kappa_E (L-s')]\right)}
{\kappa_E \xi \left({\rm Ai}[\kappa_E L]{\rm Bi'}[0]
-{\rm Ai'}[0]{\rm Bi}[\kappa_E L]\right)}~.
\label{ecorr}
\ee
\end{widetext}
Eq.(\ref{ecorr}) is the solution for $ s \geq s'$, while the
case $s<s'$ follows from symmetry.

The relevant, electric-field dependent length scale is given by
$\kappa_E^{-1}$, where \be \kappa_E= (\beta \lambda E/ \xi)^{\frac{1}{3}} ~.
\label{kappaE}
\ee It is interesting to note that this length scales as $E^{-1/3}$, which is
different than the $-1/2$ power that characterizes the dependence of
$\kappa_F^{-1}$ on the force. This could not have been predicted by dimensional
arguments. In fact, since $E\lambda L $ is a force, one might have guessed the
same scaling for $\kappa_E^{-1} $ as for $\kappa_F^{-1}$.

We can now calculate the relative extension of the polymer in the direction of
the field using Eq.(\ref{x(u)}):
\be
\frac{X}{L} =  1-\frac{\left(d-1\right)
\left({\rm Ai'}[\kappa_E L]{\rm Bi'}[0] -{\rm Ai'}[0]{\rm Bi'}[\kappa_E L]\right)}
{2 \kappa_E^2\xi L \left({\rm Ai}[\kappa_E L]{\rm Bi'}[0] -{\rm Ai'}[0]{\rm Bi}[\kappa_E L]\right)}
\label{X} ~.
\ee Eq.(\ref{X}) is one of the main new results of this paper. It provides, for
the first time, an exact expression for the field-extension relation for a
charged semi-flexible polymer in an electric field.

Since the quadratic Hamiltonian, Eq.(\ref{hquadratic}), implicitly
assumes that the fluctuations of the tangent vector are small, we
require $\avg{t_i(s)^2}$ to be small for our analysis to be self
consistent. Once again we impose the self-consistency condition,
Eq.(\ref{cond}), on the variance of the fluctuations,
$\avg{t_i(s)^2}$. It is clear from the boundary conditions, that the
tangent vector is fixed at the $s=0$ end and completely free at the
$s=L$ end, that the maximum fluctuations of the tangent vector
occurs  at $s=L$. Thus, the self-consistency requirement can be
written as \be \avg{t_i(L)^2} =\frac{{\rm Ai}[0]{\rm Bi}[\kappa_E L]
-{\rm Ai}[\kappa_E L]{\rm Bi}[0]} {\phi^\frac13\left({\rm
Ai}[\kappa_E L]{\rm Bi'}[0] -{\rm Ai'}[0]{\rm Bi}[\kappa_E
L]\right)} \leq \frac1{2(d-1)} ~. \label{condition} \ee When the
argument of Airy function is large, $Ai$ converges to $0$ whereas
$Bi$ diverges to $+\infty$ exponentially as \bea {\rm Ai}[z]
&\simeq& \frac{\exp[-\frac23 z^\frac32]}
{2\sqrt{\pi}z^\frac14} \nn \\
{\rm Bi}[z] &\simeq& \frac{\exp[\frac23 z^\frac32]}{\sqrt{\pi}z^\frac14} ~.
\label{assymptotic_Airy}
\eea
Thus, when $\kappa_E L \gg 1 $,
the condition in Eq.(\ref{condition}) reduces to
\be
\avg{t_i(L)^2} \simeq -\displaystyle
\frac{{\rm Ai}[0]}{\xi \kappa_E {\rm Ai'}[0]}
\leq \frac1{2(d-1)}
\label{large_a}
\ee
Again we get a ratio of length scales, i.e. the strong field condition
is equivalent to
\be
\kappa_E \xi \gsim 2.74 (d-1) ~,
\label{ebigE}
\ee where we have used ${\rm Ai}[0]/{\rm Ai'}[0] \approx 1.37$. The small
fluctuation approximation therefore requires that the length-scale,
$\kappa_E^{-1}$ be less than about a third of the persistence length.

On the other hand, when $\kappa_E L \ll 1$, we can Taylor expand
$Ai[\kappa_E L]$ around $0$ as ${\rm Ai}[\kappa_E L]\simeq {\rm
Ai}[0] + \kappa_E L {\rm Ai'}[0]$ and similarly for $Bi$. Then,
Eq.(\ref{condition}) becomes \be \frac{L}{\xi} \leq
\frac{1}{2(d-1)}~. \label{small_a} \ee Thus Eq.(\ref{X}) is valid
for all field strengths for polymers less than $\xi/2 $ in two
dimensions, and $\xi/4 $ in three dimensions. For longer polymers,
stronger fields are needed for its applicability. For example,
Eq.(\ref{X}) is applicable to a molecule of actin that is about $15
\mu m $ in size, and is being stretched by an electric field of at
least $0.02 \rm{V/cm}$ in strength. But it is not useful for a
molecule of DNA longer than about $25 \rm{nm}$; the minimum field
required for a $50 \rm{nm}$ DNA molecule for the small fluctuation
approximation to be appropriate is of the order of $10^2 \rm{V/cm}$.
The inset of Fig.~\ref{EField} shows the region of parameter space,
where in $d=2 $ Eq.(\ref{condition}) holds for the whole range of
$\kappa_E L$.

When the chain is long and the applied field is strong such that $E
\gg (\beta \lambda \xi^2)^{-1}$, the expression of extension reduces
to \be \frac{X}{L}=1-\frac{d-1}{2}\sqrt{\frac{k_B T}{\lambda E \xi
L}} ~. \label{bigE} \ee It should be noted that it is possible to
derive this equation using approximate phenomenological
arguments~\cite{Marko1995c, Maier2002}. A comparison with the
constant force case here is very interesting. As we have shown, the
length scale, $\kappa_F^{-1}$ in the constant force case scales as
$F^{-1/2}$ in contrast with $\kappa_E^{-1} $ that scales as
$E^{-1/3}$. In the high field limit however, the dependence of the
relative extension on the electric field becomes $~E^{-1/2}$, just
as the dependence of extension on force in the high force limit.

In the other limit, when the polymer is short so that $L/\xi \leq 1/(2d-2)$,
and the electric field is small, $E \ll (\beta \lambda \xi^2)^{-1}$,
Eq.(\ref{X}) yields the limiting linear response, \be \frac{X}{L} =
1-\frac{d-1}{4}\left(\frac{L}{\xi}\right) +\frac{(d-1)\lambda E \xi^2}{40 k_B
T}\left(\frac{L}{\xi}\right)^4 ~.
\label{smallE}
\ee Eq.(\ref{smallE}) has been derived  for $d=3 $ using different methods in
Ref.~\cite{Benetatos2004}.

\begin{figure}
\begin{center}
\includegraphics[width=4in,height=!]{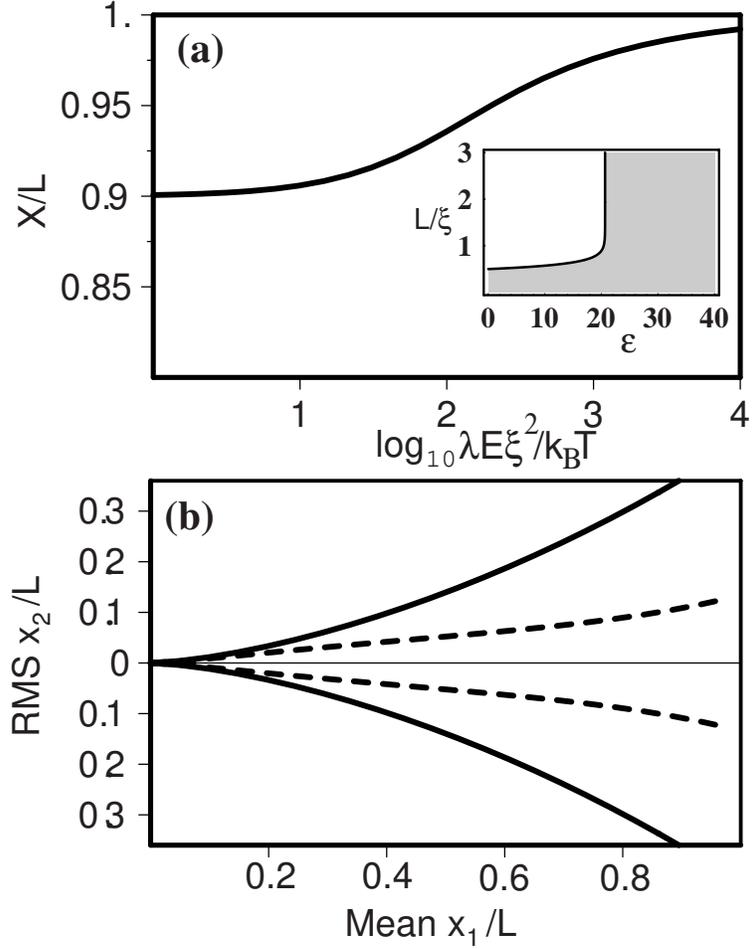}
\end{center}
\caption{ (a) Relative extension, $X/L$ is plotted against the reduced
electric field, $\beta \lambda E \xi^2$ for a polymer of size
$L/\xi=0.4 $ for $d=2$. Inset shows parameter
values allowed by self-consistency (shaded area). For actin, if
$\lambda \approx 1 \rm{e/nm}$, $\beta \lambda E \xi^2 = 1$ is about
$10^{-3} \rm{V/cm}$, for DNA with $0.6 \rm{e/nm}$ it is about $170 \rm{V/cm}$.
Note that we are assuming that the screened charge density is about one-tenth of
the bare charge density. This is consistent with experiments on
DNA~\cite{Maier2002}.
(b) The shape of a short polymer stretched by
a constant electric field in two dimensions is plotted parametrically in units
of polymer contour length $L/\xi=0.4$. The solid
curve is for $\beta \lambda E \xi^2 = 1$, and the dotted curve is for $\beta \lambda E \xi^2 = 1000$.}
\label{EField}
\end{figure}

Again we can compute the average shape, defined as before by Eq.(\ref{x(u)})
and Eq.(\ref{y(u)}), of the polymer under stretch.  In experiments on long DNA
molecules, it has been reported that the polymer assumes a trumpet-like shape,
reflecting the fact that force is stronger towards the grafted end and tends to
vanish at the free end~\cite{Maier2002}.  The integral in Eq.(\ref{x(u)}) can
be calculated analytically, while Eq.(\ref{y(u)}) needs to be calculated
numerically. A trumpet-like shape emerges naturally as the result of our
computation, and is shown in Fig.~\ref{EField}.

In experiments that stretch single molecules by electric fields, the
polymer molecule, is in an ionic solution, and an electric field
will drive a net current~\cite{Marko1995c}. The current creates a
flow in the system, and the net deformation of the molecule, it has
been argued, is therefore a combination of the flow and the
field~\cite{Stigter1998,Marko1995c,Ferree2003}. In experiments
performed with long DNA molecules of different lengths, it has been
shown that Eq.(\ref{bigE}) fits the data only with different
effective charge densities for different lengths of the molecule,
thus questioning whether the effective charge density is a
physically meaningful parameter~\cite{Ferree2003}. However we feel
that interpretation of the results have been hampered by the lack of
precise formulas for the electric field induced stretching case.
While this continues to be the case for long DNA molecules,
Eq.(\ref{X}) provides a means of testing these questions using actin
molecules. Quantitative comparison of our results with experiments
would be relatively easy for actin polymers, since the required
electric fields for the fluctuating-rod assumptions to hold are
small, and easily produced in the laboratory. Assuming that charge
screening reduces bare actin charge to about $1\,
\rm{electrons/nm}$, our results are valid for all actin lengths at
electric field strengths starting from $0.03 \,\rm{V/cm}$ upwards.
From force-extension measurements in such experiments, the effective
charge density of actin of different lengths and under different
ionic conditions can be extracted.

In view of the non-uniform nature of the effective force a charged polymer
experiences in a constant electric field, it may be useful to ask what is the
effective constant force, which would lead to the same relative extension as a
given electric field strength. Eq.(\ref{bigE}) appears to suggest that at high
forces the polymer behaves as if it is being stretched by a constant force
equivalent to that produced by one-fourth of the total charge concentrated at
the end of the chain. It turns out that over most of the regime of
applicability of our result, that is a reasonable approximation, as
demonstrated in Fig.\ref{ECompare}.

\begin{figure}
\begin{center}
\includegraphics[width=4in ,height=!]{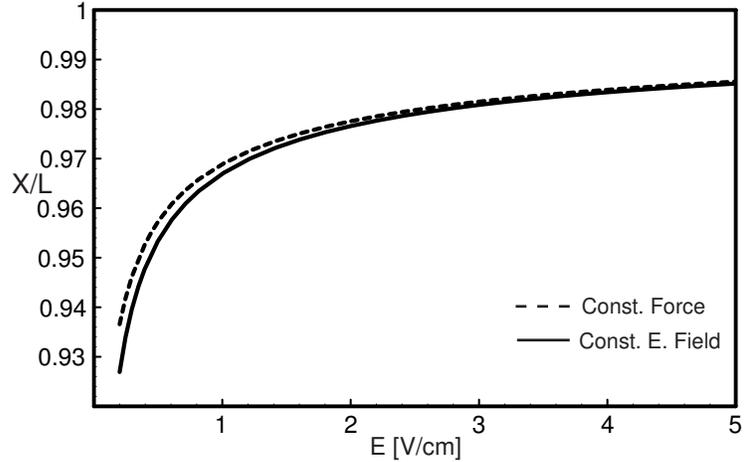}
\end{center}
\caption{\label{ECompare}The relative extension of a $15$ micron actin molecule
in a constant electric field compared to the case when it is stretched by a constant force due to a single charge on the free tip of the polymer. When this charge is chosen to be exactly $1/4$ of the total charge on the polymer, the two curves agree well over most of the range of applicability.}
\end{figure}

\section{Conclusions}
\label{secV}
Biopolymers like DNA are fast emerging as tools -- rulers and templates -- for
a number of biophysical applications, such as measuring DNA-protein
interactions. Some of these uses are due to the remarkable fact that a simple
physical model, the worm-like chain, explains DNA entropic elasticity with
great accuracy, and small deviations from this behavior, say due to bound
proteins, can be detected. However entropic effects are subtle and may be
affected by experimental conditions in ways that are not obvious. For example
it was recently pointed out that tethered bead experiments, that use a short
DNA strand attached to a bead, need to take into account entropic exclusion
forces of the bead with the wall~\cite{Segall2006}. Similarly, experiments
involving entropic elasticity need to account for the influence of experimental
conditions on the entropy of the molecule, that may quantitatively change the
force-extension relation.

In this paper we have theoretically examined entropic effects of boundary
conditions on force-extension curves. We  demonstrated that the force-extension
response of fluctuating rods is  qualitatively different from that of long
polymers, for which the contour length is much greater than its persistence
length. We also derived analytic expressions for the force-extension relation
and root-mean-square fluctuations of such a polymer stretched by a constant
force. Finally, we showed that in the fluctuating rod limit the constraint of
axis-clamping, which might be imposed by single-molecule stretching techniques
such as laser tweezers, can lead to measurable effect on force-extension
curves. More importantly, blindly employing the long-polymer formulas to data
obtained for short chains might lead to erroneous conclusions \cite{Li2006}.

Single molecule experiments using electric fields to stretch
molecules may also become a useful tool, but progress is hampered by
the lack of analytic expressions for the force-extension curves in
this case. We examine the case of a polymer stretched by an electric
field, and show that it behaves as if it is being stretched by a
non-uniform force, which increases linearly from the free end of the
chain. We derive an analytic expression for the field-extension
relation and average shape of the polymer stretched by an electric
field, in the fluctuating rod limit.

In all cases considered here, the force extension formulae,
Eq.(\ref{extlaser}) and Eq.(\ref{X}), are functions of three length
scales, two of which are the contour length $L $ and the
$3d$-persistence length $\xi $. For stretching by a constant force,
the third length scale is $\kappa_F^{-1}\equiv \sqrt{k_BT\xi/F}$,
while for a constant electric field it is $\kappa_E^{-1} \equiv
(k_BT\xi/E\lambda)^{\frac{1}{3}}$.  It is noteworthy that these
length scales do not follow from dimensional arguments alone due to
the presence of dimensionless combinations of polymer parameters.
Since these length scales are related to the decay of fluctuations
in the directions transverse to the force, and become smaller as the
polymer becomes more straight, they may be interpreted as
confinement length-scales, pointing towards deeper analogies between
stretched and confined polymers~\cite{DeGennes1979} that warrant
further exploration.

\section{Acknowledgements}
This work was supported by the NSF through grant DMR-0403997. JK is a Cottrell
Scholar of Research Corporation. We thank Paul Wiggins, Meredith Betterton, and
an anonymous referee for valuable comments.


\end{document}